**SEPnet: A sustainable model for a collaborative physics network**

Jacob Dunningham[1], Olivia Keenan[2], Cristobel Soares[3], Veronica Benson[3], Michelle Limbert[4] and Claire Hepwood[5]

1. Department of Physics and Astronomy, University of Sussex, Brighton BN1 9QH, UK
Email: executivedirector@sepnet.ac.uk
2. School of Physical and Chemical Sciences, Queen Mary University of London, London, E1 4NS, UK
3. Department of Physics, Faculty of Engineering & Physical Sciences, University of Surrey, Guildford, GU2 7XH, UK
4. School of Physics and Astronomy, University of Southampton, Highfield, Southampton, SO17 1BJ, UK
5. Department of Physics, Royal Holloway University of London, Egham Hill, Egham, Surrey TW20 0EX, UK

**Abstract**
The South East Physics network (SEPnet) is a collaboration between nine universities in the South East of England, working together to deliver excellence in physics. By sharing resources, we are able to add much more value to our departments, students, and subject than we could achieve individually. Our core ambitions include ensuring the sustainability of physics in our region, enhancing the employability of our students, delivering advanced training, securing the pipeline of future students, opening up new research opportunities and breaking down barriers to improve the accessibility of physics for everyone. We believe that SEPnet provides a tried and tested model that could be rolled out by others to improve the academic excellence of other disciplines in different regions.

## 1. Introduction

SEPnet was set up in 2008 as a collaborative network between universities in the South East of England. Its aim was to nurture physics departments at a time when low recruitment of undergraduate students threatened their financial sustainability. In its original phase, SEPnet was funded by a grant from the Higher Education Funding Council for England (HEFCE)[1] for a five-year term and supported six founding members: the universities of Kent Southampton, Surrey and Sussex as well as Queen Mary and Royal Holloway – both constituent members of the University of London; the University of Portsmouth joined in 2010. Such was the success of this initial phase that SEPnet was renewed for a further five-year term with two additional members – the Open University and the University of Hertfordshire – and much less reliance on external funding. The current third phase (SEPnet III) runs from 2018-2023 and has fulfilled the original vision of being fully sustained by contributions from the member departments who recognise the considerable benefits that membership brings.

[1] In April 2018, the duties of HEFCE were taken over by the newly-formed body United Kingdom Research and Innovation (UKRI).

Over the past 14 years the aims of SEPnet have continued to develop, driven by the evolving priorities of the members and changes to the higher education landscape in the United Kingdom. The governance structure of SEPnet has allowed for this flexibility and has been a feature of its success and longevity. SEPnet's programme for delivery is currently arranged into three core pillars: Employer Engagement, Outreach & Public Engagement and our postgraduate training programme (GRADnet). Details of these are outlined below. Underpinning each pillar is a focus on equality, diversity and inclusion (EDI). We also seek out ways we can foster research collaborations and work together on influencing policy and optimising the outcomes for all partners in government-initiated performance metrics.

**2. Employer Engagement**

In 2016, the UK Government published a review into graduate employability in Science, Technology, Engineering and Maths (STEM) subjects [1]. One of its findings was that STEM disciplines are important for economic growth and productivity, but to fully leverage this we must make sure that STEM graduates are equipped with the right skills for successful graduate employment. Recommendations included providing students with work placements, professional skills training, and increased awareness of career options, as well as creating stronger collaboration between higher education providers and employers to better align the supply and demand for STEM skills.

SEPnet addresses many of these recommendations through its Employer Engagement programme. This programme is run by a centrally-funded director who leads a team of employability advisors at each partner institution. Together they have brought together a wide network of industry contacts who employ physics graduates. SEPnet collaborates with them and shares good practice in relation to our mutual priorities. These include embedding employability skills in undergraduate courses, developing stronger links between companies and early career researchers especially through placements, arranging networking events and talks from employers, and delivering workshops.

Our core objectives are:

- Early career planning and engagement with industry for all physics students to increase the number of students going into graduate level jobs by equipping them with employability skills. We enable this by:
    - Embedding employability into the physics curriculum at our partner institutions. This includes sessions on investigating career options, CV writing and interview skills.
    - Employer talks, workshops and visits to increase our students' awareness of the career opportunities open to them. This includes running SEPnet-wide webinars on careers in different sectors.
    - A summer placement programme, which involves students being paid for eight weeks to work on a STEM project with an employer. We run about 100 such placements each year with the cost shared between employers and SEPnet partners. Case studies of these can be found here [2].
    - An annual SEPnet Expo in central London where we bring employers and students who have completed placements together to share ideas and network.
    - Expanding our network of employers and running a SEPnet employer mentoring scheme.

- Creating a pipeline of postdoctoral graduates with the employability skills needed to play a significant role in providing leadership and entrepreneurship and in driving innovation in industry. We achieve this with:
    - Industry placements, especially for final-year postgraduate research (PGR) students and with small companies and start-ups so they get a chance to develop entrepreneurship skills.
    - Maintaining and building resources such as a LinkedIn group for PGRs, alumni and employers.
    - Collecting and monitoring PGR destination data.
    - Disseminating and sharing good practice with the wider higher education community.
- Employer Engagement & Outreach working together to raise awareness of careers with physics for students at all stages. This includes:
    - Sharing careers information resources for schools and undergraduates on the SEPnet website.
    - The creation of a set of careers postcards for use with schools, to enable teachers to embed physics careers into science lessons .
    - Sharing employer and alumni contacts for public engagement events and placements.
    - Supporting employers in their outreach work with schools.
    - Building a network of employer outreach contacts to share good practice providing information on alternative career routes to physics

Each year we collate data on graduate outcomes for SEPnet institutions and compare this with the broader UK market for students in physics, maths and STEM subjects more generally. We also work closely with the White Rose Industrial Physics Academy (WRIPA) to run joint events and study the impact of, for example, students' willingness to move locations for employment [3] and how the COVID pandemic has affected the job market [4]. We continue to investigate how to close the gap between the skills graduates have and the skills employers want [5,6] and our future plans include carrying out a detailed analysis of the effectiveness of the Employer Engagement programme. Examples are already emerging that suggest it is making a real difference, such as two of our partners coming first and third for physics employability in the UK in a survey by *The Guardian* newspaper in 2021 [7], with 97% and 96% of graduates respectively in graduate-level jobs or further higher education within 15 months of graduation.

## 3. Outreach and Public Engagement (OPE)

The Outreach and Public Engagement programme is important for maintaining the pipeline of physics students for the future and keeping the public informed about physics and the research we carry out in our region. It plays a key role in SEPnet's mission to remove barriers and make physics accessible to all by targeting groups that are under-represented in physics. The OPE programme is run by a director who oversees outreach officers based at each partner university.

Our Outreach programme engages with students of all ages up until they start university. The main focus is on 11-14 year olds, but starts for children as young as under-5s with early years storytelling in libraries and public places and we are currently piloting a new primary programme for Year 3 (7 & 8 year olds). These aim to prevent children from disassociating from science at an early age. Our key message is that physics (and science generally) is for everyone and we particularly address under-served audiences in physics such as girls, students from disadvantaged backgrounds and under-represented ethnic minorities. Our Outreach programme delivers the latest research to teachers across the network through varied continuing professional development (CPD) sessions and disseminates best practice amongst academics and outreach practitioners.

Our Public Engagement work aims to ensure the public are kept informed about, inspired by, and supportive of physics research. This is done through schools and events in our local community. Our varied projects aim to reach many diverse audiences. One example is the "Tactile Universe" project [8], which makes current astronomy research accessible to the blind and vision impaired community.

Our core programmes are:

- Schools' Outreach –
    - The aim is to raise the science capital of Key Stage 3 students (11-14 year-olds) in the region by communicating the messages that anyone can do physics; physics is exciting, relevant & important and goes beyond the classroom; and studying physics further broadens career possibilities.
    - This is mainly delivered through a series of *Connect Physics* workshops [9] and the *Shattering Stereotypes* project, which raises awareness of gender stereotyping in schools and how it can affect subject choice.
    - We work with local primary schools and early years' groups to ensure we are engaging with the same pupils over the course of their educational journey.
- Public Engagement with Research –
    - The aim is to cultivate and embed a culture of engaged physics research with the public across all SEPnet partners through training and supporting research groups in SEPnet partners to develop, run and evaluate impactful methods of engaging many different and diverse publics with their research throughout their research cycles. It is based around the following three themes:
      i) Local Community – encouraging SEPnet partners to engage with communities that are local to their universities.
      ii) Research in Schools – helping run projects which enable local students in schools to engage with and carry out current physics research.
      iii) Research-Led Consultancy – assisting research groups within SEPnet partners in developing, running and evaluating impactful research-based projects that engage with different publics.

A new innovation is a collaborative recruitment project. This aims to increase the number of students choosing to study physics at a SEPnet partner university through widening participation initiatives targeting non-traditional and under-represented groups. This harks back to the original aim of SEPnet of boosting undergraduate numbers to secure the financial sustainability of our physics departments. This project will help bridge the gap between school students in their final years and university undergraduates by promoting the range of opportunities SEPnet offers as a collection of physics courses with diverse specialisms, admission grades, teaching styles and locations.

**4. GRADnet**

GRADnet is the largest physics postgraduate school in England and brings together the research strengths of our partners to deliver a wide range of training in advanced physics and professional skills beyond anything that could be delivered by any partner individually. The combined academic resources of the SEPnet partners ensure that we can offer our PGR students and early career researchers a very broad programme of physics skills training spanning many sub-disciplines. The facilities available to SEPnet allow us to offer an excellent experience through residential schools, workshops, lecture series as well as through innovative use of peer learning, video-conferencing and on-line delivery.

Our professional skills training programme helps fulfil the recommendations of the Wakeham Review [1] by equipping our students with the skills needed by future employers. It aims to address the national skills gap in scientific innovation by developing STEM leaders of the future for the benefit of the wider UK economy. A secondary benefit is that funding bodies and universities set minimum levels of training that PhD students are expected to undertake. GRADnet enables our partners to easily exceed those standards.

Our core programme includes:

- A GRADnet Induction Day for all 1st Year PhD students
- Professional skills 1-2 day workshops with a Winter School, "*Moving from science into business*" and a Summer School, "*Opportunities beyond one's PhD*".
- A range of advanced physics 1-2 day workshops that leverage the research expertise at SEPnet partners.
- Coding workshops and careers panel events in collaboration with the Institute of Physics (IOP) – the UK's physics professional body.
- Student-led conferences: students have the chance to bid for SEPnet funding to organise and run conferences of their choice. They are supported by the SEPnet team throughout and benefit not just from the conferences, but the skills and experience gained in organising them.
- Online workshops on professional skills e.g. online networking.
- An annual advanced school for PhD students offered by our "New connections between experiment and theory" (NExT) institute. This gives students a strong grounding in both theoretical and experimental aspects of particle physics, using the expertise of lecturers at the SEPnet partners.

**5. Research collaborations**

The SEPnet partners' track record of working together, strong research synergies, and the advantages of geographic proximity have opened up opportunities for research collaborations that are beyond what was originally envisioned for SEPnet.
This started with the Radiation Detection Doctoral Network (RADnet), a novel collaborative doctoral programme for research and training in radiation detection and applications. The network brought together PhD researchers, academics, industrial scientists and engineers to work on projects that are directly applicable to industrial challenges. The success of RADnet led us to set up the more ambitious Small and Medium Enterprise Doctoral Training Network (SME-DTN), which was awarded co-funding from Research England and is attracting industry sponsors and will provide up to 12 highly-leveraged PhD studentships for SEPnet partners.

SEPnet also enabled the Data Intensive Science Network (DISCnet) Centre for Doctoral Training (CDT) [10] to be established in 2017. DISCnet has trained 73 PhD students and has recently been renewed, bringing a further 18 funded PhD students to our network. CDTs like DISCnet rely on strong industry engagement and innovative training that responds to industry needs. DISCnet students benefit from GRADnet's training programme and, to date, DISCnet has managed and sourced 51 placements from 40 organisations ranging from multinationals to micro-SMEs across sectors such as finance, space and health. Many of these placement hosts have been SEPnet industrial partners. DISCnet placement partners and students are also included in SEPnet's industry/student mixer events that give insight into applying for commercial positions. In the future we will seek further opportunities for research collaboration that are enabled by SEPnet, such as strategic bids for research equipment that can be shared across the network.

**6. Equality, Diversity and Inclusion (EDI)**
Removing barriers and making physics inclusive and open to everyone is a big priority for SEPnet and underpins all of our work programmes. EDI is a good example of how we can achieve more by working collaboratively than in isolation. Our EDI work includes:

- Collecting diversity data to help us evolve our programmes to ensure that they provide value to everyone in our diverse community. Where possible, partners also share departmental cohort data to enable SEPnet to better understand and eliminate barriers to participation and success in physics.
- Diversity Leads for member departments coordinating activity amongst themselves, developing and deploying a programme based on pooling of their own resources.
- The establishment of the Intersectional PhD Peer Support network, (IPPSnet). This is a volunteer-run peer-support network for graduate students at SEPnet universities that is inclusive of students from all backgrounds. IPPSnet runs outreach and networking events for physics PhD students; hosts the *Open Dialogues Across Physics and Astronomy* student conference; and co-ordinates a peer mentoring scheme for physics PhD students across SEPnet universities.
- Women in Physics events run jointly with the Institute of Physics.
- An annual SEPnet Diversity Workshop to share good practice with STEM departments across SEPnet and the wider community.

**7. Governance**
Part of SEPnet's success has been down to its governance structure, which has had the flexibility to evolve over the three phases of the project to meet the changing needs of the partners. Each of the core programmes described above is led by a director, who together form an Operations Board overseen by an executive director. This board is responsible for the day to day running of the project, ensuring that key targets are met and that everything operates within budget. The Operations Board reports to the Collaboration Board, which is responsible for setting strategy and the overall objectives of SEPnet. This board is comprised of the heads of physics at all the partner institutions, which is important because it means all the departments are invested in the success of SEPnet and can shape it to meet their collective needs rather than someone else deciding what SEPnet should be. The Collaboration Board is chaired by one of the heads of physics on a rotating basis. The Collaboration Board in turn reports to the Steering Board, which holds it to account and decides on overarching strategic and budgetary decisions. The Steering Board is made up of senior members of the partner institutions such as deans or pro-vice-chancellors along with some independent stakeholders. This membership is important in ensuring that SEPnet has high-level support and financial assurances from all the partners. The Steering Board appoints a senior and well-respected scientist as an independent chair – currently Professor Sir William Wakeham – who is able to provide experienced independent insight into the running of SEPnet and enhance its external influence and profile.

**8. Resources**

The external funding for SEPnet has reduced through its different phases and we are now at the point where SEPnet has achieved its long-term goal of being fully sustained by the contributions of the partners. The initial HEFCE funding in phase 1 was important in getting partners on board and having the financial freedom to set up the collaboration. However, partners are now convinced of the benefits and are willing to finance it themselves. While this costs them money, it is money that many would have spent anyway on local programmes and they recognise that these would not be able to deliver the quality and scope of the programmes that SEPnet can achieve by working collaboratively. The commitment from each partner consists of:

- An annual membership fee to support the staffing and delivery for the central programmes
- Employing a part-time Outreach Officer and a part-time Employability Advisor with budgets to support their activities
- Agreeing to provide funds to help finance the SEPnet summer placements.

Institutions also commit to their PGR students engaging with the GRADnet training programme. They are more than willing to do this as it proves to be an excellent way of giving students the broader skills they need for the job market as well as ensuring universities meet the PGR training requirements stipulated by many funders.

**9. The future**

The future is bright for SEPnet. As we approach the end of the third phase of the project, we look forward to extending for a fourth term (SEPnet IV) starting in 2023. In the 14 years since its inception, SEPnet has developed a broad and expanding portfolio of activities that are of mutual benefit to the partners and is truly greater than the sum of the parts. We are pleased that our vision of a network that is self-sustained by the members' direct and in-kind contributions has been borne out. This collaboration has been successful despite the fact that market forces and government reforms of higher education in the UK pit the partners against each other as competitors for market share of students and funding.

Key to the longevity of SEPnet is our willingness to keep evolving to meet our members' changing needs and the changing higher education landscape. For SEPnet IV we plan to find ways of adding value to other subjects at our member universities. This is in response to changing administrative structures at our partners as well as a recognition of the benefits

SEPnet has brought to physics and a desire to extend this to other disciplines. Our future plans also involve finding new synergies between the different pillars such as including careers in our outreach work, and expanding our EDI programme. We hope to continue to work together delivering excellence in physics across our region for as long as our partners find it useful. We would encourage other regions and subjects to consider a similar model for sharing resources in a collaborative network that is of mutual benefit to everyone and greatly increases the opportunities and experiences for students.

**Biographies:**

Jacob Dunningham is Executive Director of SEPnet and Professor of Physics at the University of Sussex. He completed his undergraduate studies at the University of Auckland and his doctorate in the theory of Bose-Einstein condensates at the University of Oxford. He has held three personal research fellowships and faculty positions at the universities of Leeds and Sussex, where he is currently Director of Research. He is a fellow of the Institute of Physics.

Olivia Keenan is Director of Outreach and Public Engagement at SEPnet. She has an MPhys from the University of Southampton and a PhD in Astrophysics from the University of Cardiff, studying extragalactic environments. She was heavily involved in outreach and public engagement activities throughout her studies and, on completing her PhD, became the Regional Manager for London and the South East at the Institute of Physics, where she managed engagement across the region.

Cristobel Soares is the GRADnet Manager. She has been involved in SEPnet from its inception and has managed the GRADnet training programme since 2013. She previously worked as a personal assistant to senior management at Hoechst UK (now Sanofi-Aventis), Granada Television and DHL. She joined the University of Surrey in 2006 as a PA in the Centre for Environmental Strategy and in 2008 became the PA to the Head of the Department of Physics.

Veronica Benson was the Employer Liaison Director at SEPnet up until April 2022. She has a degree in Modern Languages from University of East Anglia and postgraduate qualifications in education, careers guidance and organisational behaviour. She has been the director of a conference company, and set up an education charity and alumni association. More recently she worked as South East Project Manager for WISE (Women In Science and Engineering) and led a national STEM project on the benefits of placements for undergraduates.

Michelle Limbert is co-Employer Liaison Director at SEPnet. She has an LLB Hons (Law and French) degree from Manchester Metropolitan University, a Masters in European and International Law from the Université de Toulouse and a Certificate in Political Science from Science Po Aix. She has worked with the European Commission supporting UK industry abroad and, more recently, for the University of Southampton focusing on employer engagement, knowledge transfer and promoting entrepreneurship.

Claire Hepwood is co-Employer Liaison Director at SEPnet. She has a BSc in Computer Science with modern languages and has held a number of positions in the high performance computing industry ranging from Software Analyst to Professional Services Consultant. While at AWE, she took on the position of Strategic Outreach Scientist working for the Chief Scientist and was influential in AWE becoming an active member of WISE.